# Ghost imaging without beam splitter

Shibei He, Xia Shen, Hui Wang, Wenlin Gong and Shengsheng Han

*Key Laboratory for Quantum Optics and Center for Cold Atom Physics, Shanghai Institute of Optics and Fine Mechanics, Chinese Academy of Sciences, Shanghai 201800, China*

Compiled July 6, 2009

Many significant results have been achieved in the fields of ghost imaging, in which the beam splitter is an indispensable optical component. This paper introduces a method to realize ghost imaging without beam splitter. And we study this method experimentally and theoretically. Finally, we suggest that our device can be applied to implement the ghost imaging when we use the Sun light as the light source.

Ghost imaging has attracted much attention since it was realized initially by entangled photon pairs[1]. In the mid2000s, pseudo-thermal radiation is proposed to be utilized as lamp house to produce the same phenomenon, and then the experiment is implemented[2][3][4].

In former experiments, a beam splitter is always a necessary element. It is set in the pseudo-thermal light field, and produces two identical light fields, both in near field and far field. One of them goes through an optical system comprising an object, and then is recorded by a detector. This optical path is named "object arm", or "test arm". The other passes another optical system named "reference arm" which doesn't embody the object, and finally is recorded by another detector. The data of the two probers is correlated to yield the image or the Fourier transformation image of the object[5].

However, when we want to use the Sun light as the light source, the device's limitation emerges, for it is very unpractical to place a BS in the optical path.

The effect of BS is making two identical light fields, physically speaking, creating both near and far field correlations. On this condition, two arms' optical axes experience the same optical field, though they are separated in space. If we remove BS, the two arms' optical axes must have an angle, coinciding on the same origin on the ground glass, corresponding to two different momentum directions $k$ and $k'$, which possess no correlation at all. This is the sticking point why BS is indispensable and Sun light ghost imaging without beam splitter has little progress yet.

Some effort has been done to remove BS. In[6], a grating is in place of BS, and the Fourier transformation image is obtained. From the physical viewpoint, the grating creates a pure far-field correlation. But this experiment still needs a light-splitting instrument. Thus utilizing pure near-field correlation is considered in this paper.

In our experiment, we use an improved pseudo-thermal radiation and a special "reference arm" to obtain the Fourier transformation image of the object.[**Fig1**]

The former pseudo-thermal light consists of a laser beam radiating on a rotating ground glass. The effect of the ground glass is giving various parts of the beam's cross section a random phase modulation. This device was invented to simulate the statistical properties of real thermal light[7]. The real thermal light field has two basic characters. One is violate random fluctuation both in luminous plane and the field outside, the other is that the field obeys the Gaussian statistics[8]. The pseudo-thermal light has every property except one. Because the intensity of the laser beam's cross section is uniform, there is no intensity fluctuation on the luminous plane and in the very near area around the ground glass. So, the pseudo-thermal light is an incomplete simulation of the real thermal light. The core of our improvement is generating the intensity modulation on the rough surface of the ground glass, making non uniform intensity distribution on the luminous plane.

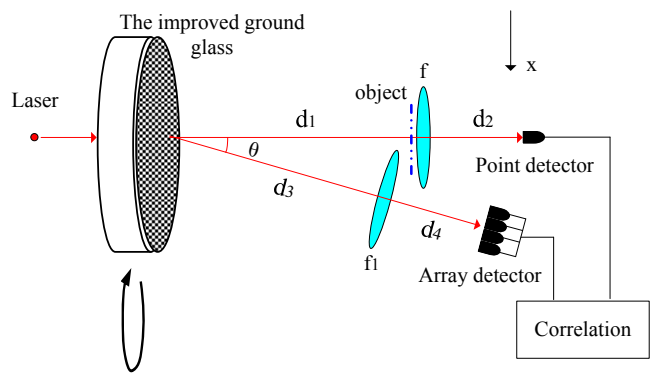

Fig. 1. The setup of the experiment

The "object arm" is placed vertical to the ground glass. It consists of an object placed after the light source with distance $d_1$, and a lens(focus=$f$), immediately after it, and finally a point detector(test detector)set after the lens with distance $d_2$. Because there is no space between the object and the lens, we assume the two are on the same plane. $d_1$ and $d_2$ should satisfy the relationship: $\frac{1}{d_1} + \frac{1}{d_2} = \frac{1}{f}$.

The "reference arm" is put near the "object arm". The optical axis has a small angle $\theta$ to that of the "object arm". And the two axes have the same origin on the ground glass. A lens(focus=$f_1$)is placed from the origin with distance $d_3$, vertical to the axis. After that, an array detector is set with the distance $d_4$. The Gaussian thin lens formula $\frac{1}{d_3} + \frac{1}{d_4} = \frac{1}{f_1}$ should be satisfied.

For simplicity, only one dimension $x$ is considered. The intensity modulation on the ground glass doesn't change the statistical properties of the light field, so the field is a complex circular Gaussian process[8]. Under this condition, the correlation between the intensity fluctuations at the test detector(object arm) and the reference detector(reference arm) is:

$$\langle \Delta I_t(x_t) \Delta I_r(x_r) \rangle =$$
$$|\int dx_1 dx_1' G^{(1,1)}(x_1, x_1') h_t(x_1, x_t) h_r(x_1', x_r)|^2 \quad (1)$$

In this equation, $h_t(x_1, x_t)$ is the impulse response function of the object arm(test arm), $x_1$ is the coordinates on the light source plane, and $x_t$ is on the test detector's plane. Similarly, $h_r(x_1', x_r)$ is the impulse response function of the reference arm. $x_1'$ is the coordinates on the light source plane, and $x_r$ is on the plane of the reference detector. $G^{(1,1)}(x_1, x_1')$ is the second-order correlation function on the light source plane. $G^{(1,1)}(x_1, x_1') = \langle E(x_1)^* E(x_1') \rangle$[8]. Suppose the laser beam's intensity distribution is uniform before the ground glass, and enough data acquisition is taken, each point's ensemble average intensity value is equal on the luminous plane.

$$G^{(1,1)}(x_1, x_1') = I_0 \delta(x_1 - x_1') \quad (2)$$

According to Fourier optics[8], if a light travels through a free space $d$, from plane $x_1$ to $x_2$, the impulse response function is:

$$h(x_1, x_2) = \sqrt{\frac{e^{ikd}}{i\lambda d}} exp[\frac{i\pi}{\lambda d}(x_1 - x_2)^2] \quad (3)$$

In this equation, $k = \frac{2\pi}{\lambda}$, $\lambda$ is the wavelength. If a light passes a lens (focus=$f$, and the lens is on the plane of $x$), the light field will be added a quadratic phase factor $exp[\frac{-i\pi}{\lambda f} x^2]$.

Firstly we consider the object arm(test arm). The beam from the ground glass travels through sequentially a free space $d_1$, object $t(x_2)$, immediately a lens (focus=$f$), a free space $d_2$, and is finally received by the test detector. So, according to Fourier optics, we get the $h_t(x_1, x_t)$:

$$h_t(x_1, x_t) = \int dx_2 \sqrt{\frac{e^{ikd_1}}{i\lambda d_1}} exp[\frac{i\pi}{\lambda d_1}(x_1 - x_2)^2]$$
$$t(x_2) exp[\frac{-i\pi}{\lambda f} x_2^2] \sqrt{\frac{e^{ikd_2}}{i\lambda d_2}} exp[\frac{i\pi}{\lambda d_2}(x_2 - x_t)^2] \quad (4)$$

In the equation, $x_2$ is coordinates on the plane of the object and the lens $f$.

According to Gaussian thin lens formula, the reference arm makes the image of the intensity modulation of the ground glass on the array detector plane. Because the reference arm's optical axis is not vertical to the ground glass, the object is not the exact intensity modulation, but its projection on the plane vertical to the optical axis. So the size of the image is the contracted size of the exact intensity modulation with the proportion $\cos\theta$. Then $h_r(x_r, x_1')$ is

$$h_r(x_1', x_r,) = \delta(d_4 \cos\theta x_1' + d_3 x_r) exp[ig(x_r)] \quad (5)$$

$exp[ig(x_r)]$ is a phase factor, and it represents every point's phase on the image. In the derivation afterwards, we can see this phase factor is of no importance.

Substituting Equ(2), (4), (5) to Equ(1), we have:

$$\langle \Delta I_t(x_t) \Delta I_r(x_r) \rangle = |\int dx_1' \int dx_1 \int dx_2 I_0 \delta(x_1 - x_1')$$
$$\sqrt{\frac{e^{ikd_1}}{i\lambda d_1}} exp[\frac{i\pi}{\lambda d_1}(x_1 - x_2)^2] t(x_2) exp[\frac{-i\pi}{\lambda f} x_2^2] \sqrt{\frac{e^{ikd_2}}{i\lambda d_2}}$$
$$exp[\frac{i\pi}{\lambda d_2}(x_2 - x_t)^2] \delta(d_4 \cos\theta x_1' + d_3 x_r) exp[ig(x_r)]|^2 \quad (6)$$

On the assumption of the infinity of the light source and the lens, we have:

$$\langle \Delta I_t(x_t) \Delta I_r(x_r) \rangle = |\frac{I_0}{\sqrt{-\lambda^2 d_1 d_2}} exp[\frac{ikd_1 + ikd_2}{2} + ig(x_r)$$
$$+ \frac{i\pi}{\lambda d_1}(\frac{x_r d_3}{d_4 \cos\theta})^2 + \frac{i\pi}{\lambda d_2} x_t^2] \int dx_2 exp[(\frac{i\pi}{\lambda d_1} \frac{2 x_r d_3}{d_4 \cos\theta}$$
$$+ \frac{-i 2 x_t \pi}{\lambda d_2}) x_2] t(x_2)|^2$$
$$= |\frac{I_0}{\sqrt{-\lambda^2 d_1 d_2}} \int dx_2 exp[(\frac{i\pi}{\lambda d_1} \frac{2 x_r d_3}{d_4 \cos\theta} + \frac{-i 2 x_t \pi}{\lambda d_2}) x_2] t(x_2)|^2$$
$$= \frac{I_0^2}{\lambda^2 d_1 d_2} |T(\frac{x_r d_3}{\lambda d_1 d_4 \cos\theta} - \frac{x_t}{\lambda d_2})|^2 \quad (7)$$

$T$ is the Fourier transformation function of $T$. If we choose a fixed point on the test arm, for instance $x_t$=0, the Fourier transformation image is achieved when we scan the reference arm $\langle \Delta I_t(x_t) \Delta I_r(x_r) \rangle = \frac{I_0^2}{\lambda^2 d_1 d_2} |T(\frac{x_r d_3}{\lambda d_1 d_4 \cos\theta})|^2$. The same image as the above can be obtained when we use a lens(focus=$\frac{2 d_1 d_4 \cos\theta}{d_3}$, which we call " effective focus") in the $f - f$ system. We can validate our experiment via this method[9].

In our experiment, the object we choose is a double slit. The width of the each slit is $60\mu m$, and the interval between the centers of the two slits is $120\mu m$. Other parameters are as follows: $f = 7.5cm$, $d_1 = d_2 = 15.0cm$, $f_1 = 10.0cm$, $d_3 = d_4 = 20.0cm$, and $\theta$ is measured $20.7°$ So the effective focus is $15 \times \cos\theta = 14.0cm$. [**Fig2**] is the result.

We obtain the zeroth and the first diffraction spectrum. Because the intensity modulation's size isn't small enough, the resolution ratio is relatively low. Due to that, the peaks in [b] and [c] don't separate distinctly. In [c], the calculated distance between the main peak and the side peak is 110 pixels. So the corresponding value of our experiment should be $110 \times \cos\theta = 103$ pixels, but in [b]

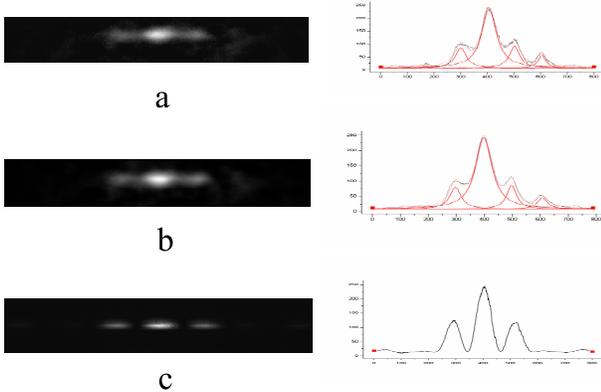

Fig. 2. [a] is the single-point correlation image, [b] is the multi-point correlations' average image, c is the standard Fourier transformation image when we use a lens(focus=15cm) in the $f-f$ system. The right are the plots corresponding to the left. The red lines in the figure[a] and figure[b] are Multi-peak fitting results.

the experimental one after the Multi-peak fitting process is 101 pixels. The error is 2%.

The mechanism of ghost imaging is two-photon interference, from the macroscopic view, the far and near field correlation[10]. If neither of them exist, no image can be achieved. The entangled photon pairs, possess both the far field and the near field correlation. That is, to have $\Delta(p_1 + p_2) = 0$ and $\Delta(x_1 - x_2)$ simultaneously [11][12]. However, the pseudo-thermal light (the setup of laser beam radiating on the ground glass) has neither of them, so intensity modulation on luminous plane is necessary.

Implementing ghost imaging with Sun light is an exciting topic at present. Our work is a significant progress to the realization of this conceit. Utilizing the self-intensity distribution on the luminous plane of the Sun light, we can only place a lens and a point detector after the object. And we can get the Fourier Transformation of the object by observing the light source via the reference arm. But there is still a problem, that we can't always let the object and our "imaging lens" immediately contacted. But through optical field transformation, we can easily overcome this difficulty.

In conclusion, we devise a method to realize ghost imaging without beam splitter, and study it theoretically and experimentally. Physically speaking. It is the near-field correlation on the plane of the light source that really functions. Finally our experiment setup is potential to be applied in the ghost imaging with the Sun light.